\documentclass[onecolumn,preprintnumbers,amsmath,amssymb]{revtex4}
\usepackage{amsmath, amsthm, amssymb,graphicx,mathtools,algpseudocode,algorithmicx,bbm,hyperref}
\usepackage[latin1]{inputenc}
\begin{document}
\title{An exact solution to Brownian dynamics of a reversible bimolecular reaction in one dimension}
\author{Stephen Smith}
\affiliation{School of Biological Sciences, University of Edinburgh, Mayfield Road, Edinburgh EH9 3JR, Scotland, UK}

\author{Ramon Grima}
\affiliation{School of Biological Sciences, University of Edinburgh, Mayfield Road, Edinburgh EH9 3JR, Scotland, UK}
 \begin{abstract}
Brownian dynamics is a popular fine-grained method for simulating systems of interacting particles, such as chemical reactions. Though the method is simple to simulate, it is generally assumed that the dynamics is impossible to solve exactly and analytically, aside from some trivial systems. We here give the first exact analytical solution to a non-trivial Brownian dynamics system: the reaction $A+B\xrightleftharpoons[]{}C$ in equilibrium in one-dimensional periodic space. The solution is a function of the particles' diffusion coefficients, radii, length of space and unbinding distance.
\end{abstract}
\maketitle
Assuming that physical quantities follow Brownian motion is a popular method of simplifying complex behaviour. In finance option prices are assumed to locally follow Brownian motion \cite{black1973pricing}, in ecology animals are assumed to perform roughly Brownian random walks \cite{bartumeus2005animal}, and in chemistry solute particles are assumed to follow Brownian paths through a solvent \cite{gillespie2012simple}. In each case, the true underlying dynamics of the system (involving people's investment decisions, animals' response to the environment, the position and momentum of each solvent particle respectively) is far too complicated to be studied in its entirety, and is instead assumed to be well-approximated by a sequence of random perturbations. 

Although the Brownian assumption makes computer simulations of these systems considerably faster, it is nonetheless commonly assumed to be analytically intractable in all but the most trivial cases. In particular in chemical kinetics, several authors have proposed numerical techniques for Brownian dynamics simulations of chemical reactions such as the exact simulation Green's function reaction dynamics (GFRD) method \cite{van2005simulating}, and the approximate but fast method \emph{Smoldyn} \cite{andrews2004stochastic}. On the other hand, analytical results about the random dynamics of solute particles are extremely rare \cite{lipkova2011analysis} .

In this letter, we consider the simplest non-trivial Brownian dynamics system: $A+B \xrightleftharpoons[k_2]{k_1}C$.  Specifically, we consider a system on a periodic line segment, which initially consists of a single particle $C$ diffusing in space. This particle can unbind with rate $k_2$ to produce two particles $A$ and $B$ which themselves diffuse on the periodic line segment until they collide and become $C$ again. The particles obey Smoluchowski dynamics, in that they react immediately upon collision \cite{smoluchowski1917physik}: this implies that the effective binding rate $k_1$ is entirely determined by the diffusion of the particles and cannot be chosen in advance.  

\begin{figure}
\includegraphics[trim=0.8cm 14cm 6cm 0cm, clip=true,scale=0.55]{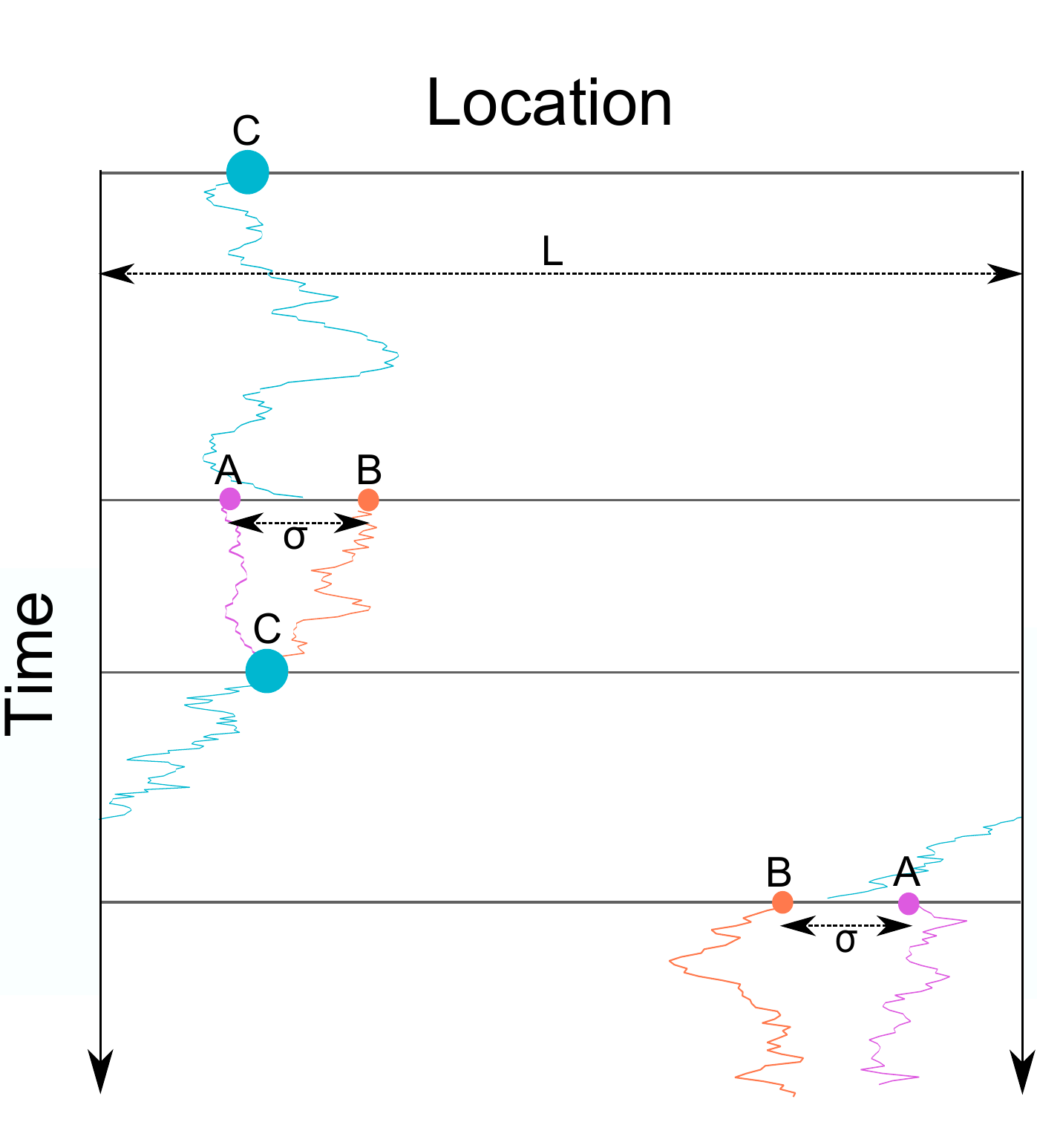}
\caption{An example of the system we study in this letter. A single $C$ particle (blue) diffuses on a line segment of length $L$ until it unbinds to form an $A$ (purple) and a $B$ (orange) particle separated by a distance $\sigma$. These diffuse on the line until they collide and become a $C$ particle. The process then repeats indefinitely. Periodic boundaries ensure that particles can diffuse over the edge of the line and appear on the other side.}\label{fig1}
\end{figure}

An example of this system is shown in Fig. \ref{fig1}, in which $C$ is represent by a blue circle, and $A$ and $B$ by purple and orange circles respectively. Note that the periodic line segment implies that particles can diffuse over one edge and appear on the other. The collision aspect of this example system is related to the one-dimensional trapping reaction problem, in which a large number point particles diffuse on an infinite line and annihilate upon collision with each other \cite{redner2001guide}. Determining the asymptotic long-time behaviour of such a system is a popular problem, though clearly very different from the problem we tackle here.

Statistically, we can think of our problem as a system with two alternating states: bound ($C$) and unbound ($A$ and $B$). We are interested in the equilibrium probability of being in the bound state, $P_\text{bound}$, or equivalently the probability of being in the unbound state $P_\text{unbound}=1-P_\text{bound}$. The absolute location of the particles is of little relevance to this problem, since the periodic line segment has no boundaries. When the system is in the bound state, we can observe the system from the frame of reference of $C$, such that nothing happens until the unbinding event. When the system is in the unbound state, we can observe the system from the frame of reference of $B$, such that the only motion is the movement of $A$ relative to $B$. This simplification allows us to focus only on the times between binding and unbinding.

The dynamics of the system are the binding and unbinding events, which occur alternately and at random times. Unbinding events are classically modelled as Poisson processes \cite{van1992stochastic}, thus times of unbinding events are assumed to follow an exponential distribution with a mean equal to the reciprocal of the reaction rate: we will use this assumption here. After unbinding, the $A$ and $B$ particles are placed an unbinding distance $\sigma$ away from each other. The binding events of our system, however, cannot be assumed to follow any standard distribution, since they are entirely dependent on the time taken for the two particles $A$ and $B$ to collide. For now, we simply say that the time until a binding event after unbinding follows a \emph{collision} distribution (which we assume has a finite mean), which we will investigate in more detail later. Note that since the system is initially in the bound state, each binding event will be independent and identically distributed (i.i.d.), since they all start with $A$ at a fixed unbinding distance from $B$.

Let the periodic line segment have length $L$, let $A$ and $B$ have radii $r_A$ and $r_B$ and diffusion coefficients $D_A$ and $D_B$ respectively, let the unbinding distance be $\sigma>r_A+r_B$ (otherwise the particles will immediately react), and let $X_i$ be i.i.d. collision-distributed random variables (the time until a collision) and $Y_i$ be i.i.d. exponentially distributed random variables (the time until unbinding). By the ergodic property, the equilibrium probability of being in the bound state is given by the proportion of time spent in the bound state over a long trajectory. In other words:
\begin{equation}
P_\text{bound}=\text{lim}_{n \rightarrow \infty} \frac{Y_1+...+Y_n}{X_1+...+X_n+Y_1+...+Y_n}.
\end{equation}
Now consider the sequences $x_n=\frac{X_1+...+X_n}{n}$ and $y_n=\frac{Y_1+...+Y_n}{n}$. By definition, $P_\text{bound}=\text{lim}_{n \rightarrow \infty} \frac{y_n}{x_n+y_n}$. By the weak law of large numbers, $x_n$ converges in probability to a constant $\mu_X$, where $\mu_X$ is the mean of the collision distribution. Similarly, $y_n$ converges in probability to $\mu_Y$, the mean of the exponential distribution. By Slutsky's theorem \cite{slutsky1925stochastische,gut2012probability}, $x_n+y_n$ converges in probability to $\mu_X+\mu_Y$. Again, by Slutsky's theorem, $\frac{y_n}{x_n+y_n}$ converges in probability to $\frac{\mu_Y}{\mu_X+\mu_Y}$. It follows that:
\begin{equation}\label{eq2}
P_\text{bound}= \frac{\mu_Y}{\mu_X+\mu_Y}.
\end{equation}
The constant $\mu_Y$ is given immediately by the unbinding rate: $\frac{1}{k_2}$. It therefore remains to evaluate $\mu_X$.

From the frame of reference of $B$, $A$ will intially be located at $\sigma$, and will subsequently diffuse with the combined diffusion coefficient $D=D_A+D_B$, while $B$ remains stationary at $0$. The collision will occur when $A$ first hits one of the point $r_A+r_B$ and $L-r_A-r_B$, i.e. when it first leaves the interval $[r_A+r_B,L-r_A-r_B]$. The probability of $A$ being located at a point $x$ at time $t$ after unbinding, $p(x,t)$, is then the solution of a diffusion equation with absorbing boundaries:
\begin{align}\label{diffeq}
\frac{\partial}{\partial t} p(x,t)=D\frac{\partial^2}{\partial x^2}p(x,t),~p(r_A+r_B,t)=p(L-r_A-r_B,t)=0,~p(x,0)=\delta(x-\sigma).
\end{align}
The survival probability of $A$ is then simply given by $P_s(t)=\int_{r_A+r_B}^{L-r_A-r_B}p(x,t) dx$. The probability that the reaction has occured by time $t$ is then $1-P_s(t)$. This is the cumulative distribution function of the first passage time distribution of the collision event, hence the quantity $\mu_X$ is then the expected value of this distribution, in other words:
\begin{equation}\label{eq4}
\mu_X=-\int_{0}^\infty t P_s'(t) dt=-\int_{0}^\infty t \frac{\partial}{\partial t} \int_{r_A+r_B}^{L-r_A-r_B}p(x,t) dx dt,
\end{equation}
Where -$P_s'(t)=\frac{1-\partial P_s(t)}{\partial t}$ is the probability density function of the first passage time distirbution. Integration by parts implies that Eq. \eqref{eq4} can be written simply as:
\begin{equation}\label{mu}
\mu_X=\int_{0}^\infty  \int_{r_A+r_B}^{L-r_A-r_B}p(x,t) dx dt,
\end{equation}
so it remains to find $p(x,t)$. For simplicity, we change the variables of Eq. \eqref{diffeq} to $x'=x-r_A-rB$, so that the diffusion equation takes place in the range $[0,l]$ where $l=L-2r_A-2r_B$ and with $\sigma^\star=\sigma-r_A-r_B$. 
\begin{align}\label{diffeq2}
&\frac{\partial}{\partial t} q(x',t)=D\frac{\partial^2}{\partial x'^2}q(x',t),~q(0,t)=q(l,t)=0,~q(x,0)=\delta(x'-\sigma^\star).
\end{align}We make the ansatz:
\begin{align}
q(x',t)=&A_0(t)+\sum_{n=1}^\infty \left[ A_n(t) \text{cos}\left(\frac{n \pi x'}{l}\right)+B_n(t) \text{sin}\left(\frac{n \pi x'}{l}\right)\right].
\end{align}
This ansatz solves Eq. \eqref{diffeq2} if:
\begin{align}
A_0(t)&=a_0,\nonumber\\
A_n(t)&=a_ne^{-D\left( \frac{n \pi}{l}\right)^2 t},\\
B_n(t)&=b_ne^{-D\left( \frac{n \pi}{l}\right)^2 t}\nonumber.
\end{align}
The left boundary condition states that:
\begin{equation}
a_0+\sum_{n=1}^\infty a_ne^{-D\left( \frac{n \pi}{l}\right)^2 t}=0 ~\forall t.
\end{equation}
The only way the sum can be zero for all $t$ is if $a_n=0$ for all $n=0,1,2,...$. The same is implied by the right boundary condition. It follows that $q(x',t)=\sum_{n=1}^\infty b_ne^{-D\left( \frac{n \pi}{l}\right)^2 t}\text{sin}\left(\frac{n \pi x'}{l}\right)$. The initial distribution can be written as a Fourier sine series:
\begin{equation}
\delta(x'-\sigma^\star)=\sum_{n=1}^\infty \frac{2}{l} \text{sin} \left( \frac{n \pi \sigma^\star}{l}\right)\text{sin} \left(\frac{n \pi x'}{l}\right),~~x\in [0,l],
\end{equation}
which allows us to identify the values of the $b_n$'s. It follows that solution of Eq. \eqref{diffeq2} is:
\begin{equation}
q(x',t)=\sum_{n=1}^\infty \frac{2}{l} \text{sin} \left( \frac{n \pi \sigma^\star}{l}\right)\text{sin} \left(\frac{n \pi x'}{l}\right)e^{-D\left(\frac{n \pi}{L}\right)^2t},
\end{equation}
which we can convert back to the original notation to obtain the solution of Eq. \eqref{diffeq}:
\begin{align}\label{diffeqsol}
p(x,t)=\sum_{n=1}^\infty \frac{2}{L-2r_A-2r_B} \text{sin} \left( \frac{n \pi (\sigma-r_A-r_B)}{L-2r_A-2r_B}\right)
 \text{sin} \left(\frac{n \pi (x-r_A-r_B)}{L-2r_A-2r_B}\right)e^{-D\left( \frac{n \pi}{L-2r_A-2r_B}\right)^2 t}.
\end{align}
Applying Eq. \eqref{mu} to this expression gives:
\begin{align}\label{mu2}
\mu_X=-\sum_{n=1}^\infty \frac{2((-1)^n-1)(L-2r_A-2r_B)^2}{(D_A+D_B)n^3\pi^3} \text{sin}\left(\frac{n \pi (\sigma-r_A-r_B)}{L-2r_A-2r_B} \right)=\frac{g(r_A,r_B,L,\sigma)}{D_A+D_B},
\end{align}
where, $z=\frac{\sigma-r_A-r_B}{L-2r_A-2r_B}$, 
\begin{align}g(r_A,r_B,L,&\sigma)=\frac{i (L-2r_A-2r_B)^2}{\pi^3}\left[Li_3\left(-e^{i\pi z}\right) -Li_3\left(e^{i\pi z}\right) -Li_3\left(-e^{-i\pi z}\right)+Li_3\left(e^{-i\pi z}\right) \right],
\end{align} and $Li_3$ is the third polylogarithm function. Using this value of $\mu_X$, we can write an expression for the effective binding rate $k_1$:
\begin{equation}\label{eq16}
k_1=\frac{1}{\mu_X}=\frac{D_A+D_B}{g(r_A,r_B,L,\sigma)}.
\end{equation}
Remarkably, this is directly proportional to the combined diffusion coefficient $D_A+D_B$. This implies that a fast diffusion limit of Smoluchowski Brownian dynamics is absurd: particles will react immediately after unbinding. We also have an expression for the equilibrium probability of being in the bound state $P_\text{bound}$ obtained using Eq. \eqref{eq2}, Eq. \eqref{mu2}, and the fact that $\mu_Y=\frac{1}{k_2}$:
\begin{equation}\label{pbound}
P_\text{bound}=\frac{D_A+D_B}{D_A+D_B+k_2g(r_A,r_B,L,\sigma)}.
\end{equation}
The equilibrium probability of being in the bound state is therefore a Michaelis-Menten-like function of the combined diffusion coefficient $D_A+D_B$, and the corresponding Michaelis-Menten constant is directly proportional to the unbinding rate $k_2$. 
\begin{figure}
\includegraphics[scale=0.4]{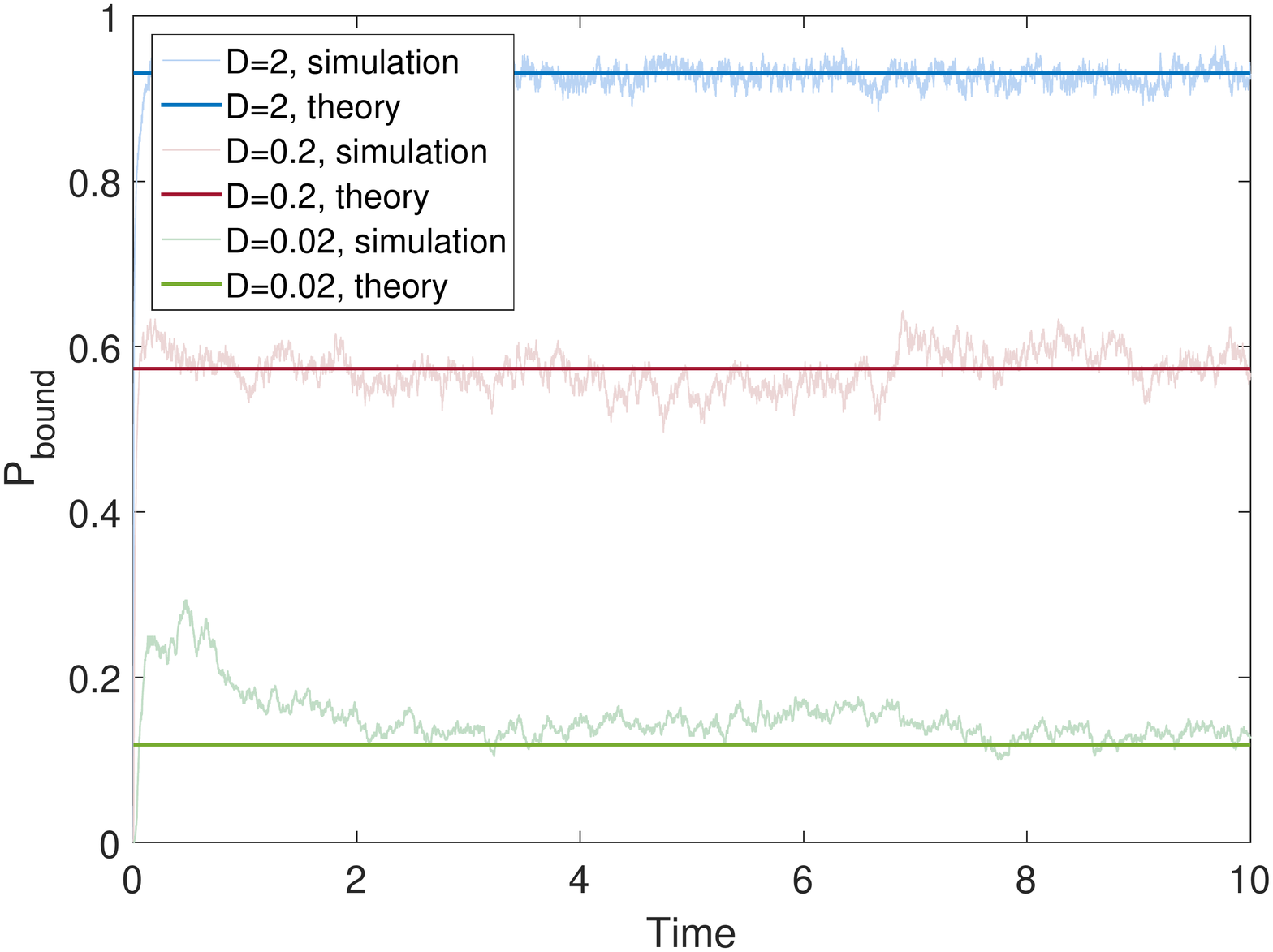}
\caption{Comparison of Eq. \eqref{pbound} (solid lines) with BD simulations of the system described in Fig. \ref{fig1} (fluctuating lines), for three different values of the combined diffusion coefficient $D=D_A+D_B$. The simulations are an ensemble average of 500 trajectories computed with an Euler-Maruyama scheme. Parameters are  $r_A=r_B=0.02$, $\sigma=0.11$, $k_2=5$, $L=1$, $\Delta t=10^{-5}$.}\label{fig2}
\end{figure}
To confirm our theory, we performed computer simulations of the example system described in Fig. \ref{fig1}. We used an Euler-Maruyama scheme which updated $A$ particle positions by a Normal$(0,\sqrt{2 D_A \Delta t})$ random number and $B$ positions by a Normal$(0,\sqrt{2 D_B \Delta t})$ random number at each time step of length $\Delta t$. A reaction was deemed to occur as soon as the distance between the centres of $A$ and $B$ was less than $r_A+r_B$, using the periodic distance metric $d(x,y)=\text{min} \lbrace \vert x-y \vert, L-\vert x-y \vert \rbrace$. For the unbinding reaction, an exponential random number was chosen by which to advance the time and the particles were placed a distance $\sigma$ apart. 

In Fig. \ref{fig2} we plot an average of 500 trajectories of the simulation described above, for three different values of the combined diffusion coefficient $D=D_A+D_B$. The thick solid lines show the value predicted by Eq. \eqref{pbound} for each parameter set. It is clear that our formula predicts the correct equilibrium behaviour for each diffusion coefficient. This is therefore an independent verification of our results.

Our results can be applied to investigate the reaction system on an infinite line segment. Taking a large $L$ expansion of the function $g$ gives:
\begin{equation}
g(r_A,r_B,L,\sigma)= \frac{\sigma-r_A-r_B}{2}L+O(1).
\end{equation}
It follows that the bimolecular reaction rate for large $L$ is given by:
\begin{equation}\label{eq19}
k_1=\frac{2(D_A+D_B)}{(\sigma-r_A-r_B)L}.
\end{equation}
On an infinite line segment, therefore, the bimolecular reaction will, on average, never happen, and the system will remain in the unbound state indefinitely. Note that in practice it may be more practical to use Eq. \eqref{eq19} than Eq. \eqref{eq16}, if $L$ is large.

In conclusion, we have shown that (i) it is possible to derive an exact analytical expression for a non-trivial Brownian dynamics system involving collisions between diffusing particles, (ii) the effective collision rate between two Brownian particles is a linear function of their diffusion coefficients, (iii) the equilibrium probability distribution of the system $A+B \xrightleftharpoons[]{}C$ has a Michaelis-Menten-like formula, which depends on diffusion coefficients, particle radii, length of space, and unbinding distance, (iv) on an infinite line segment the system has a trivial equilibrium distribution which is in the unbound state with probaiblity $1$. Note that our result also solves the system $A+A \xrightleftharpoons[]{}B$, because nothing prevents $A$ and $B$ from having the same radii and diffusion coefficients. 

Our results hints that further analytical results about more complex Brownian dynamics systems may be possible. Going beyond 1 dimension is a significant challenge, but we have derived approximate (not exact) analytical expressions for the reaction $A+B \xrightleftharpoons[]{} C$ in 2 and 3 dimensions, which will be the subject of a future paper.

\acknowledgments
This work was supported by a BBSRC EASTBIO PhD studentship to S.S. and by a Leverhulme grant award to R.G. (RPG-2013-171).
\bibliography{mybibfile}{}

\end{document}